\documentclass[twocolumn, twocolappendix]{aastex631}

\usepackage{comment}
\usepackage{amsmath}
\usepackage{graphicx}
\usepackage[caption=false]{subfig}
\begin{document}

\title{Monster radio jet ($>$66 kpc) observed in quasar at $z\sim5$ }

\author[0009-0009-8274-441X]{Anniek J. Gloudemans}
\affiliation{NSF NOIRLab, Gemini Observatory, 
670 N A'ohoku Place
Hilo, HI 96720, USA}

\author[0000-0002-6470-7967]{Frits Sweijen}
\affiliation{Centre for Extragalactic Astronomy, Department of Physics, Durham University, Durham DH1 3LE, UK}

\author[0000-0003-0487-6651]{Leah K. Morabito}
\affiliation{Centre for Extragalactic Astronomy, Department of Physics, Durham University, Durham DH1 3LE, UK}
\affiliation{Institute for Computational Cosmology, Department of Physics, Durham University, Durham DH1 3LE, UK }

\author[0000-0002-6822-2254]{Emanuele Paolo Farina}
\affiliation{International Gemini Observatory/NSF NOIRLab, 670 N A’ohoku Place, Hilo, Hawai'i 96720, USA}

\author[0000-0001-6889-8388]{Kenneth J. Duncan}
\affiliation{Institute for Astronomy, Royal Observatory, Blackford Hill, Edinburgh, EH9 3HJ, UK}

\author[0000-0002-6047-430X]{Yuichi Harikane}
\affiliation{Institute for Cosmic Ray Research, the University of Tokyo, 5-1-5 Kashiwa-no-Ha, Kashiwa City, Chiba, 277-8582, Japan}

\author[0000-0001-8887-2257]{Huub~J.~A. R\"{o}ttgering}
\affiliation{Leiden Observatory, Leiden University, PO Box 9513, 2300 RA Leiden, The Netherlands}

\author[0000-0001-5333-9970]{Aayush Saxena}
\affiliation{Department of Physics, University of Oxford, Denys Wilkinson Building, Keble Road, Oxford OX1 3RH, UK}

\author[0000-0002-4544-8242]{Jan-Torge Schindler}
\affiliation{Hamburger Sternwarte, University of Hamburg, Gojenbergsweg 112, D-21029 Hamburg, Germany}

\begin{abstract} 

We present the discovery of a large extended radio jet associated with the extremely radio-loud quasar J1601+3102 at $z\sim5$ from sub-arcsecond resolution imaging at 144 MHz with the LOFAR International Telescope. These large radio lobes have been argued to remain elusive at $z>4$ due to energy losses in the synchrotron emitting plasma as a result of scattering of the strong CMB at these high redshifts. Nonetheless, the 0.3\arcsec\ resolution radio image of J1601+3102 reveals a Northern and Southern radio lobe located at 9 and 57 kpc from the optical quasar, respectively. The measured jet size of 66 kpc makes J1601+3102 the largest extended radio jet at $z>4$ to date. However, it is expected to have an even larger physical size in reality due to projection effects brought about by the viewing angle. Furthermore, we observe the rest-frame UV spectrum of J1601+3102 with Gemini/GNIRS to examine its black hole properties, which results in a mass of 4.5$\times$10$^{8}$ M$_{\odot}$ with an Eddington luminosity ratio of 0.45. The BH mass is relatively low compared to the known high-$z$ quasar population, which suggests that a high BH mass is not strictly necessary to generate a powerful jet. This discovery of the first $\sim100$ kpc radio jet at $z>4$ shows that these objects exist despite energy losses from Inverse Compton scattering and can put invaluable constraints on the formation of the first radio-loud sources in the early Universe. 
\end{abstract}


\section{Introduction} \label{sec:intro} 

Despite the discovery of radio-loud\footnote{Quasars are classified as radio-loud when the radio-loudness $R>10$, defined as $R = F_{5\text{GHz}}/F_{4400\text{\AA}}$ in rest-frame \citep{Kellermann1989AJ.....98.1195K}.} quasars and radio galaxies up to $z
\sim7$ (e.g. \citealt{McGreer2006ApJ...652..157M, Willott2010AJ....139..906W, Banados2015ApJ...804..118B, Saxena2019MNRAS.489.5053S, Belladitta2020A&A...635L...7B, Banados2021ApJ...909...80B, Gloudemans2022A&A...668A..27G, Endsley2023MNRAS.520.4609E}), there appears to be a lack of large ($\sim$100s kpc) radio lobes at $z>4$ with the most extended radio jet measured to be 36 kpc at $z=4.1$ (e.g. \citealt{DeBreuck1999A&A...352L..51D, Saxena2024MNRAS.531.4391S}) and 1.6 kpc at $z\sim6$ \citep{Momjian2018ApJ...861...86M}. The lack of extended radio sources above $z>4$ has previously been attributed to the CMB energy density, which increases with (1+$z$)$^4$ and causes energy losses of relativistic electrons in the radio jet by inverse Compton (IC) scattering \citep{Fabian2014MNRAS.442L..81F, Ghisellini2014MNRAS.438.2694G}. This effect causes high redshift extended jets to become X-ray bright and radio weak. However, even with the most powerful X-ray telescopes, such as Chandra, it has been challenging to observe extended X-ray jets at $z>4$ (see e.g. \citealt{Connor2021ApJ...911..120C, Ighina2022A&A...659A..93I}). 

The new generation of powerful radio telescopes, such as the Low Frequency Array (LOFAR; \citealt{vanHaarlem2013A&A...556A...2V}) and the near future Square Kilometre Array (SKA; \citealt{Dewdney2009IEEEP..97.1482D}), enables a combination of deep and wide field imaging at frequencies of $\sim$100 MHz for the first time. By combining the international LOFAR stations a sub-arcsecond resolution (of 0.3\arcsec) can be achieved at 144 MHz (see e.g. \citealt{Varenius2015A&A...574A.114V, Ramirez-Olivencia2018A&A...610L..18R, Harris2019ApJ...873...21H, Morabito2022MNRAS.515.5758M, Sweijen2022A&A...658A...3S, Sweijen2022NatAs...6..350S}). This presents the opportunity to study distant radio sources at low frequencies in exquisite detail.  

In this Letter, we present the discovery of a large ($>66$ kpc) extended radio jet at $z>4$ for the first time using LOFAR long baseline imaging at 0.3\arcsec\ resolution. The resolved radio jet is associated with the extremely radio-loud quasar J1601+3102 at $z=4.9$ that was discovered recently by \cite{Gloudemans2022A&A...668A..27G} using the LOFAR Two Metre Sky Survey Data Release 2 (LoTSS-DR2; \citealt{Shimwell2022A&A...659A...1S}). In this work, we study its radio properties and derive its black hole properties from follow-up rest-frame UV spectroscopic observations using the Gemini Near-Infrared Spectrograph (GNIRS; \citealt{Elias2006SPIE.6269E..4CE}). The discovery of this source reveals the existence of extended radio sources into the cosmic dawn despite the increased CMB energy density. Throughout this work we assume a $\Lambda$-CDM cosmology with H$_{0}$= 70 km s$^{-1}$ Mpc$^{-1}$, $\Omega_{M}$ = 0.3, and $\Omega_{\Lambda}$ = 0.7 and use the AB magnitude system.

\section{Data}
\label{sec:data}

The high-$z$ quasar J1601+3102 was discovered as part of a sample of 20 radio-bright quasars, which were selected by combining an optical dropout selection with low-frequency radio observations (see \citealt{Gloudemans2022A&A...668A..27G} for details). J1601+3102 stood out from this sample with an exceptionally high radio luminosity and steep spectral index. Therefore, to further explore its radio jet and SMBH properties, we constructed a LOFAR VLBI image at 144 MHz and observed the quasar with Gemini/GNIRS to obtain its (near-)infrared spectrum. 

\subsection{International LOFAR Telescope data reduction}
The ILT data presented in this work was taken as part of the LOFAR Two-metre Sky Survey (PI: Shimwell, project code LT10\_010). Observations were carried out in the usual LoTSS fashion (\citealt{Shimwell2022A&A...659A...1S}) consisting of a ten-minute observation of a flux density calibrator (3C 295, L656058) followed by an eight-hour observation of the target field (P240+30, L656064). A total of $51$ stations participated ($24$ core stations, $14$ remote stations, $13$ international stations). The observation setup is summarised in Table~\ref{tab:obs_ilt}.

\begin{table}
    \centering
    \caption{ILT observation setup}
    \label{tab:obs_ilt}
    \begin{tabular}{l l l}
        \hline \hline
        Central frequency & MHz & 144 \\
        Frequency range & MHz & 120 - 168 \\
        Pointing centre & J2000 & $\alpha = 16^{\text{h}}00^{\text{m}}28^{\text{s}}.892$ \\
         & & $\delta = +30^{\circ}00'06''.911$ \\
        Integration time & s & 1 \\
        Channel width & kHz & 12.207 \\
        Observation time & s & 28,800 \\
        Target distance & deg & 1.07 \\
        Delay calibrator distance & deg & 0.72 \\
        Target-delay calibrator sep. & deg & 0.52 \\
        \hline
    \end{tabular}
\end{table}

Data processing consisted of three parts: correcting for systematics, calibrating the Dutch array and calibrating the international array. This was done using the LOFAR Initial Calibration (LINC; \citealt{deGasperin2019A&A...622A...5D}) pipeline for the systematics and for direction-independent calibration of the Dutch array. Calibration for the international array was done using the LOFAR-VLBI pipeline \citep{Morabito2022A&A...658A...1M} and \texttt{facetselfcal} \citep{vanWeeren2021A&A...651A.115V}. Calibration solutions were derived using DP3 \citep{Dijkema2023}. Imaging was done using \texttt{WSClean} \citep{Offringa2014}. Details of the full calibration procedure can be found in Appendix~\ref{appendix:LOFAR_ILT}.

\begin{figure*}
\centering 
\subfloat[]{%
  \includegraphics[width=1.0\columnwidth]{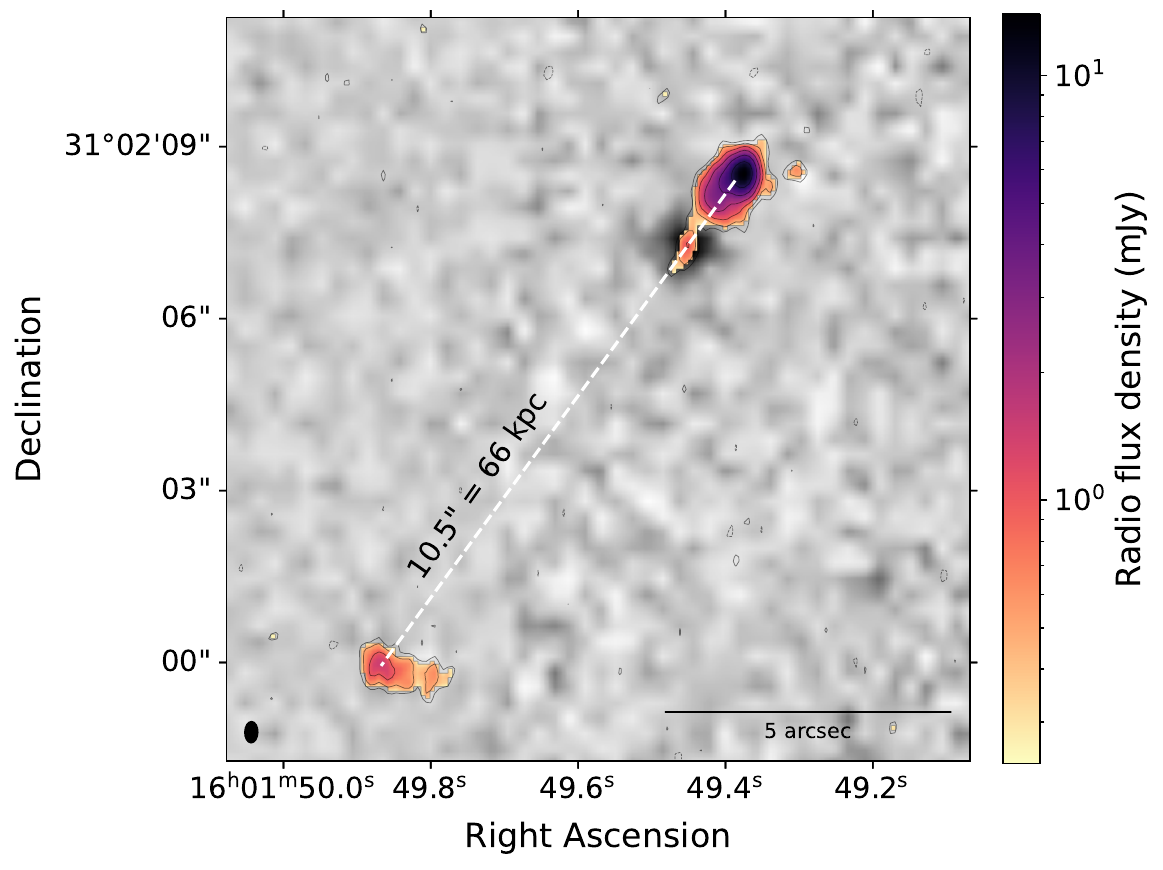}
}\qquad
\subfloat[]{%
  \includegraphics[width=\columnwidth]{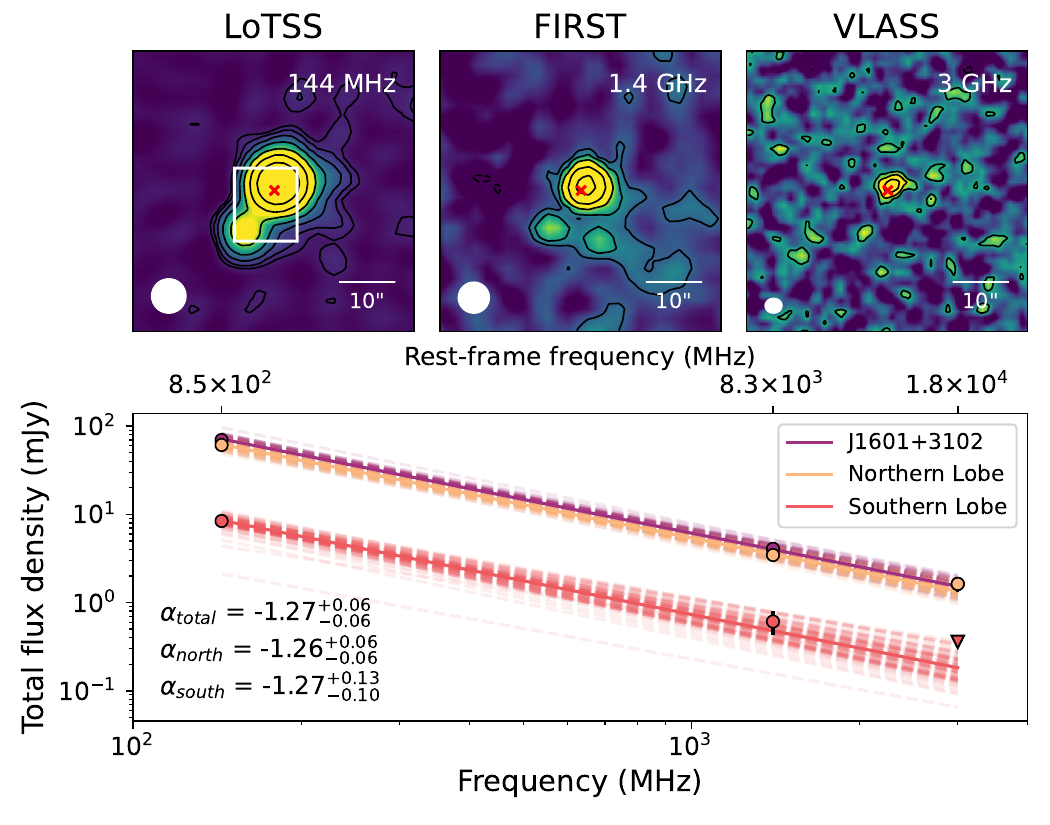}
}\vspace{-0.4cm}
\caption{\label{fig:vlbi_im} \textbf{Left:} LOFAR VLBI image of the extended radio jet of J1601+3102 at 144 MHz superimposed on an optical $z$-band image of the DESI Legacy Imaging Survey. The radio contours are drawn at [$-1$, 1, 2, 4, 8, 16, 32]$\ \times\ 3\sigma_{\text{rms}}$ with $\sigma_{\text{rms}} = 0.08$ mJy beam$^{-1}$. The beam size (resolution of 0.3\arcsec) is shown in the bottom left corner. The source shows a Northern and Southern lobe at a distance of 1.4 and 8.9\arcsec\ from the optical quasar, which equals a projected distance of 9 and 57 kpc, respectively. The physical size of the extended radio jet is therefore $>66$ kpc, making this the largest radio jet at $z>4$ to date. \textbf{Top right:} Low-resolution 50\arcsec\ radio cutouts from LoTSS, FIRST, and VLASS with radio contours drawn at [$-1$, 0.5, 1, 2, 4, 8, 16, 32]$\ \times\ 3\sigma_{\text{rms}}$. The respective beam sizes are indicated in the left bottom corner. The white rectangle corresponds to the LOFAR VLBI image in the left panel and the red crosses indicate the position of the optical quasar. \textbf{Bottom right:} Radio spectrum from 144 MHz to 3 GHz derived from these three radio surveys. The Northern lobe is detected in all surveys, however, the Southern lobe is only detected in LoTSS and FIRST. The non-detection in VLASS is given as a 3$\sigma$ upper limit. The spectra are well described by a power law with the scatter indicated by the dashed lines. The spectral indices derived from these surveys are nearly identical with $-1.26^{+0.06}_{-0.06}$ and $-1.27^{+0.13}_{-0.10}$ for the Northern and Southern lobe, respectively.}
\end{figure*}

The resulting LOFAR VLBI image ($\sigma_{\text{rms}}\sim$0.08 mJy beam$^{-1}$) is shown in Fig.~\ref{fig:vlbi_im}, which shows an extended radio jet with two lobes and core emission. We extract the source components and their flux densities using the Python Blob Detector and Source Finder (\textsc{PyBDSF}; \citealt{Mohan2015ascl.soft02007M}). Finally, to account for systematic offset in the flux calibration, a 10\% flux density scale uncertainty is added in quadrature to the resulting flux density measurements. The radio structure, source association, and radio properties are discussed in Sect.~\ref{sec:results_radio_jet}.

\subsection{Archival radio data}
\label{subsec:archival_radio_data}

To obtain the radio spectral index\footnote{Defined as $S_{\nu} = \nu^{\alpha}$} of the different components of J1601+3102, we utilize archival data from LoTSS-DR2, the VLA FIRST survey at 1.4 GHz (\citealt{Becker1994ASPC...61..165B}), and the Very Large
Array Sky Survey at 2-4 GHz (VLASS; \citealt{Lacy2020PASP..132c5001L}). The available VLASS data for our source is an Epoch 3.1 Quick Look continuum image. J1601+3102 is detected in each of the three radio surveys and their radio flux densities are again extracted using \textsc{PyBDSF} (see Sect.~\ref{sec:results_radio_jet}). 

\subsection{(Near-)infrared spectroscopic follow-up}

The infrared spectrum of J1601+3102 was obtained with Gemini-North/GNIRS on the 19th of March 2024 (GN-2024A-FT-102, PI: Gloudemans) using the cross-dispersed mode (32 l/mm). This configuration provides wavelength coverage from 0.8-2.5 $\mu$m with a spectral resolution of $R\sim650$, which allows for resolving the Mg\textsc{ii}, C\textsc{iii}] and C\textsc{iv} broad emission lines of the quasar. We observe J1601+3102 using the standard ABBA slit nodding technique (3\arcsec\ offset) with a slit width of 0.675\arcsec\ and 0.15\arcsec\ pixel scale. The total time on target was 77 mins with single exposures of 230 seconds each, resulting in five ABBA sequences. Preceding our science observation, we observe the standard star HIP73156 (type A1V, V=6.504 mag) for telluric correction and flux calibration. We reduce the data using the \textsc{python} package Python Spectroscopic Data Reduction Pipeline (\textsc{PypeIt}\footnote{\url{https://github.com/pypeit/PypeIt}}; \citealt{Prochaska2020JOSS....5.2308P}), which provides semi-automated reduction for spectroscopic data. Details of this reduction procedure are given in Appendix~\ref{appendix:NIR_data_reduction}.

\begin{figure*}
    \centering
    \includegraphics[width=\textwidth]{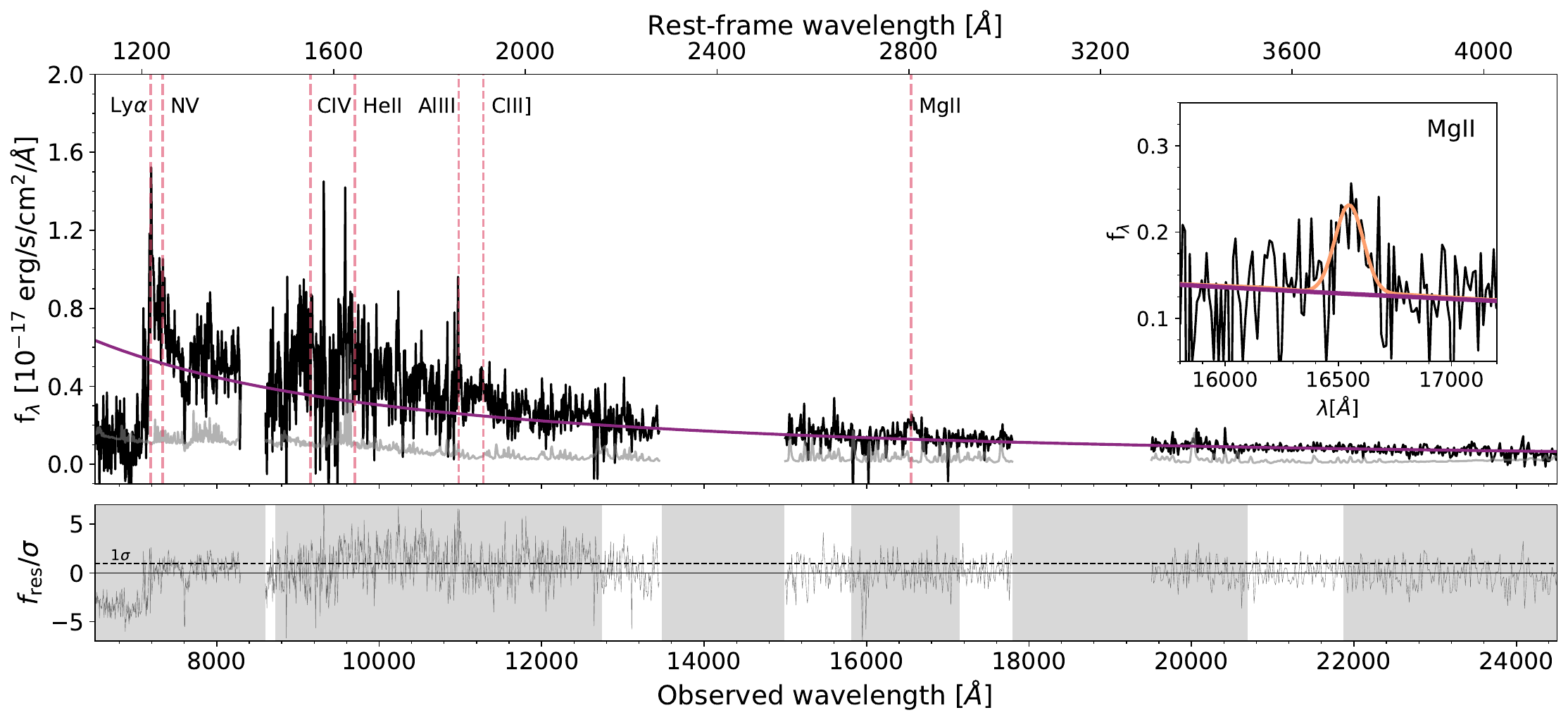}
    \caption{\textbf{Top panel:} Composite optical and infrared spectrum of J1601+3102 obtained with HET/LRS2 ($<8500$ \AA) and Gemini/GNIRS ($>8500$ \AA) binned to a resolution of 200 km s$^{-1}$. The error of the GNIRS spectrum (grey) increases significantly towards bluer wavelengths. The continuum is described by a power-law + iron pseudocontinuum (purple line; see Sect.~\ref{subsec:spectral_fitting}). The wavelength regions heavily affected by telluric lines are masked for visualization purposes. The inset shows the Mg\textsc{ii} line including the best fit with an FWHM of 2694$^{+510}_{-383}$ km s$^{-1}$. \textbf{Bottom panel:} The residual flux divided by the error spectrum after subtracting the continuum and Ly$\alpha$, N\textsc{v}, C\textsc{iv}, C$\textsc{iii]}$, and Mg\textsc{ii} emission line fits (see Appendix~\ref{appendix:spectral_fitting}). The grey shaded regions show the masked regions for continuum fitting.} 
    \label{fig:composite_spec}
\end{figure*}

We combine the reduced GNIRS spectrum with the optical spectrum obtained in \cite{Gloudemans2022A&A...668A..27G} with the Hobby Eberly Telescope LRS2 integral field spectrograph (HET/LRS2; \citealt{Ramsey1998AAS...193.1007R, Hill2021AJ....162..298H}) in Texas, USA (see \citealt{Gloudemans2022A&A...668A..27G} for details). The HET spectrum covers a wavelength range of $6450-10500 \AA$ with a spectral resolution of $R\sim1800$. We utilize the \textsc{python} package \textsc{Sculptor} \citep{Schindler2022ascl.soft02018S} to create the composite spectrum and fit the continuum and emission lines (see Sect.~\ref{subsec:spectral_fitting} for details).
We performed absolute flux calibration on both the optical and infrared spectra using the Legacy DECAM $z$-band magnitude of 21.19$\pm$0.07, since the quasar is not detected in existing wide-field near-infrared imaging surveys. We create the composite spectrum by normalizing the optical spectrum to the infrared spectrum.
Finally, we re-bin both spectra onto a common wavelength resolution of 200 km s$^{-1}$. The resulting spectrum used for analysis is shown in Fig.~\ref{fig:composite_spec} including a zoom-in on the detected Mg\textsc{ii} broad emission line. In this figure, we masked the regions heavily affected by telluric contamination for visualization purposes.

\subsection{Spectral fitting}
\label{subsec:spectral_fitting}

For our spectral fitting procedure of the rest-frame UV quasar spectrum, we again use the \textsc{Sculptor} package, considering both the continuum emission and broad emission lines. The continuum model generally consists of three components: a power-law, Balmer pseudocontinuum, and iron pseudocontinuum. This method has been widely used and detailed in previous work for modeling quasar spectra (e.g. \citealt{DeRosa2014ApJ...790..145D, Mazzucchelli2017ApJ...849...91M, Shen2019ApJ...873...35S, Schindler2020ApJ...905...51S, Banados2021ApJ...909...80B, Farina2022ApJ...941..106F}) and therefore we only provide a brief summary here.

The accretion disk emission is modeled with a normalized power-law with slope $\alpha_{\lambda}$ at a rest-frame wavelength of 2500\AA. We do not include contribution from the Balmer continuum emission, because the spectral quality varies quite a bit, especially towards the bluer wavelengths. Finally, the iron contribution from the broad-line region, which is especially prominent around Mg\textsc{ii}, is modeled using an empirical iron template of \cite{Vestergaard2001ApJS..134....1V} derived from a narrow–line Seyfert 1 galaxy I Zwicky-1. To correctly model the iron emission in our spectrum, this iron template is broadened with the FWHM of our Mg\textsc{ii} line. We fit this continuum model to wavelength regions free of emission lines, strong telluric absorption, and unusually large flux errors: $\lambda_{\text{rest}} =$ 1455-1475\AA, 2155-2280\AA, 2535-2675\AA, 2900-3010\AA, 3500-3700\AA. 

We subtract the continuum model from the spectrum before fitting the broad emission lines. We fit the Ly$\alpha$, N\textsc{v}, C\textsc{iv}, C$\textsc{iii]}$, and Mg\textsc{ii} emission lines using a single Gaussian (see Appendix~\ref{appendix:spectral_fitting}).
Because the Mg\textsc{ii} fit and the iron contribution in the continuum model depend on each other, we iteratively fit both until the FWHM of the Mg\textsc{ii} line converges.

To obtain confidence levels on these fitting parameters, we resample the observed spectrum 1000 times by assuming a Gaussian distribution of flux values centered around the measured flux value, with a standard deviation equivalent to the flux uncertainty, and refit each spectrum. The final best fit parameters are given by the median of the distribution with 1$\sigma$ uncertainties given by the 16 and 84th percentile. 

\section{Large extended radio jet}
\label{sec:results_radio_jet}

The LOFAR VLBI image of J1601+3102 reveals an extended radio structure including a Northern radio lobe, Southern radio lobe, and a core (see Fig.~\ref{fig:vlbi_im}). 
The Northern lobe is located at 1.4\arcsec\ from the optical quasar host (9 kpc projected distance) with a total flux density of 50.6$\pm$5.1 mJy. The Southern lobe is offset at 8.9\arcsec\ (57 kpc projected distance) with a total flux density of 10.5$\pm$1.6 mJy. 

\subsection{Two radio lobes}
\label{subsec:two_radio_lobes}

To investigate whether the Southern lobe is related to J1601+3102, we study the system's geometry, radio spectra of both lobes, and the probability of misassociation. As a simple test, we connect the locations of the peak flux densities of both lobes. This line runs straight through the middle of the optical quasar host as shown in the left panel of Fig.~\ref{fig:vlbi_im}, which is the first indication this could be the counter jet. In addition, the Southern radio feature has the expected curved shape of a radio lobe propagating outward. To measure the spectral indices of radio features, we make use of the low-resolution archival radio data (see Sect.~\ref{subsec:archival_radio_data}). The resulting low-frequency radio spectrum is displayed in the right panel of Fig.~\ref{fig:vlbi_im}. The Southern lobe is not significantly detected in VLASS and therefore a 3$\sigma$ upper limit is given to constrain the spectral index. The spectral slopes of both lobes are remarkably similar with slopes of $-1.26^{+0.06}_{-0.06}$ and $-1.27^{+0.13}_{-0.10}$ for the Northern and Southern lobe, respectively. The steepness and similarity of the two spectral indices are a strong indication these both originate from J1601+3102, since steep slopes are often found in lobe-dominated radio galaxies (see e.g. \citealt{Miley2008A&ARv..15...67M, Tadhunter2016A&ARv..24...10T}). The Northern lobe does not show any evidence of a high-frequency break caused by an ageing electron population in the radio lobes from the three data points (see e.g. \citealt{Jaffe1973A&A....26..423J, Murgia2003PASA...20...19M, callingham2015ApJ...809..168C}). However, this could be due to the low number of measurements.

We note there is a 3$\sigma$ detected radio source in the FIRST image of 0.62$\pm$0.22 mJy (see right panel of Fig.~\ref{fig:vlbi_im}), which is not detected in either LoTSS or VLASS. Given the local rms in those images, the spectral indices of this source are constrained to be $\alpha^{1.4\text{GHz}}_{144\text{MHz}} \geq 0.01$ and $\alpha^{3\text{GHz}}_{1.4\text{GHz}} \leq -0.7$, which indicates it is possibly a faint giga-hertz peaked spectrum (GPS) source. We suspect this is either a GPS source unrelated to J1601+3102 or a noise spike (since S/N$\lesssim3$) and therefore do not include it in further analysis.

Finally, since there is no optical counterpart in any of the $g-$, $r-$, or $z-$band images of the DESI Legacy Imaging Survey (DECaLS; \citealt{dey2019AJ....157..168D}) at the position of the Southern lobe and no infrared counterpart detected in the $W1, W2, W3,$ and $W4$ band images of the Wide-field Infrared Survey Explorer (WISE; \citealt{wright2010AJ....140.1868W}) either, we calculate the probability that this radio source is associated with an optically faint radio galaxy below our detection limit. To determine this probability, we utilize the LoTSS Deep Fields data in the ELAIS-N1 and Lockman Hole fields \citep{Kondapally2021A&A...648A...3K, tasse2021A&A...648A...1T, sabater2021A&A...648A...2S, Duncan2021A&A...648A...4D}, which both reach a sensitivity of $\sim$20 $\mu$Jy beam$^{-1}$ and have extensive multi-wavelength coverage. To calculate the number density of $>8$ mJy radio galaxies below the 5$\sigma$ detection limit of the Legacy surveys, we simply determine the number of galaxies in the multi-wavelength catalogs of these fields that meet these criteria and divide that by the total survey area of 8.05 and 13.32 deg$^2$ for ELAIS-N1 and Lockman Hole, respectively. This results in an expected number of $\sim$9$\times10^{-5}$ radio galaxies hiding within 100 arcsec$^{2}$ of our quasar. We do note that this calculation assumes a constant number density of galaxies, while quasars are known to reside in more strongly clustered environments (e.g. \citealt{Garcia2017ApJ...848....7G}). These faint optical galaxies typically have magnitudes of $\sim$23-26 in $z-$band. The possible existence of such a source could therefore be confirmed with deep imaging. 
Combining all the previous arguments, we conclude that the Southern radio feature is the counter jet of this quasar. 

\subsection{Physical jet properties}

With a radio jet size of 66 kpc, J1601+3102 is the most extended radio jet ever observed at $z>4$ (see Fig.~\ref{fig:BH_mass} right panel) with the largest previously known jets from a radio galaxy 36 kpc at $z=4.1$ (e.g. \citealt{DeBreuck1999A&A...352L..51D, Pentericci2000A&A...361L..25P, Saxena2024MNRAS.531.4391S}) and other literature high-$z$ quasars never exceeding $\sim$2 kpc (\citealt{Momjian2018ApJ...861...86M}). However, this projected jet size is only a lower limit, since its physical size is likely larger due to projection effects brought about by the viewing angle. The orientation-based unification scheme predicts that in the case of radio-loud quasars the radio axis is oriented within 45$^{\circ}$ of the observer's line of sight (e.g. \citealt{Barthel1989ApJ...336..606B, Urry1995PASP..107..803U}). Considering a viewing angle $\theta > 45^{\circ}$ as measured from the radio axis to the plane of the sky, this would imply a lower limit on the physical jet size of $>93$ kpc. 

The radio core is not resolved and has a total flux density of 2.2$\pm$0.3 mJy. We calculate the brightness temperature of the different components using,
\begin{equation}
    T_b = \Bigg(\frac{S_{\nu}}{\text{min}\times\text{maj}}\Bigg) \times \Bigg(\frac{1.22\times 10^{12}}{\nu^2}\Bigg) \times (1+z)
\end{equation}
with $S_{\nu}$ the total flux density in Jy, $\nu$ the observed frequency in GHz, and min and maj the minor and major axis in mas, respectively (see e.g. \citealt{Morabito2022MNRAS.515.5758M}). Since the core is not resolved, we use the deconvolved size measurements to set a lower limit on T$_{b}$, which yields T$_{b} > 0.86 \times 10^{7} $ K. For the Northern and Southern lobe we measure brightness temperatures of $6.3 \times 10^{7}$ K and $1.0 \times 10^{7}$ K, respectively, which is as expected above the typical AGN limit of T$_{b}\sim$10$^{5-6}$ K at 144 MHz \citep{Morabito2022MNRAS.515.5758M} and similar to the resolved radio jet measurements of \citealt{Momjian2018ApJ...861...86M} at $z\sim6$.

The radio lobes of J1601+3102 are highly asymmetric in terms of their brightness and distance from the quasar. As apparent from Fig.~\ref{fig:vlbi_im}, the Northern lobe is notably brighter ($\sim5$$\times$) and geometrically closer ($\sim$6$\times$) to the quasar. In general, the approaching hotspot is expected to be geometrically further away in the place of the sky from the quasar since that lobe has been able to grow larger before the light from the receding jet arrives. The Southern lobe is therefore likely the approaching jet, whereas the Northern lobe is the receding jet. The apparent brightening of the Northern lobe and asymmetry of the system is likely caused by the local environment, such as jet interaction with the (dense) ISM (e.g. \citealt{McCarthy1991ApJ...371..478M, Nesvadba2017A&A...599A.123N, Dutta2024arXiv240100446D}). This has been observed in other high-$z$ radio galaxies as well, such as TN J1338-1942 where the Southern lobe is about 3 times more distant than the Northern lobe and $\sim$4 times fainter \citep{Pentericci2000A&A...361L..25P}. Follow-up observations (e.g. deep imaging or Integral Field Unit) are necessary to measure the extended ionized gas around J1601+3102 and to confirm the environmental effects and jet-gas interaction that may be at play.

\section{Black hole properties}
\label{sec:bh_properties}

\begin{figure*}
\centering 
\subfloat[]{%
  \includegraphics[width=\columnwidth]{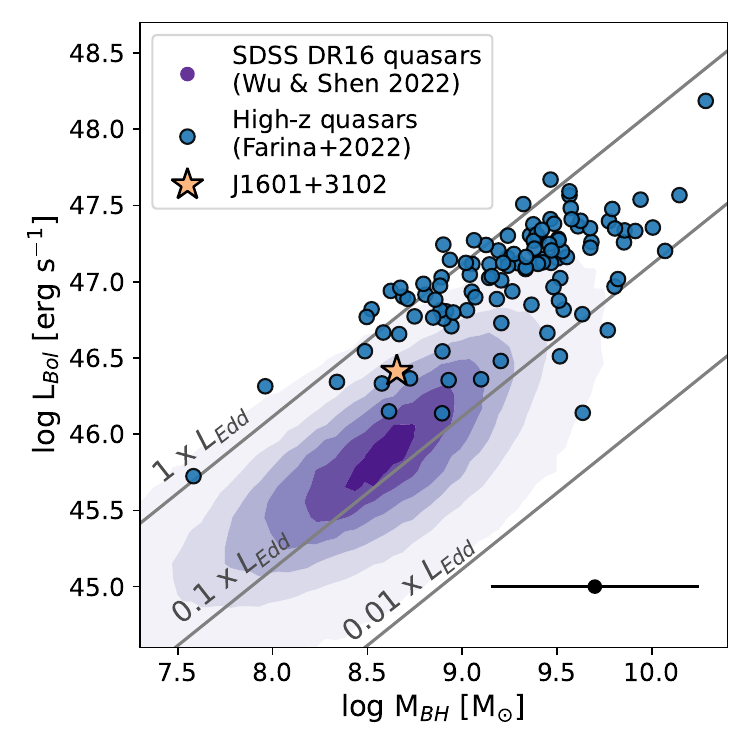}
}\qquad
\subfloat[]{%
  \includegraphics[width=\columnwidth]{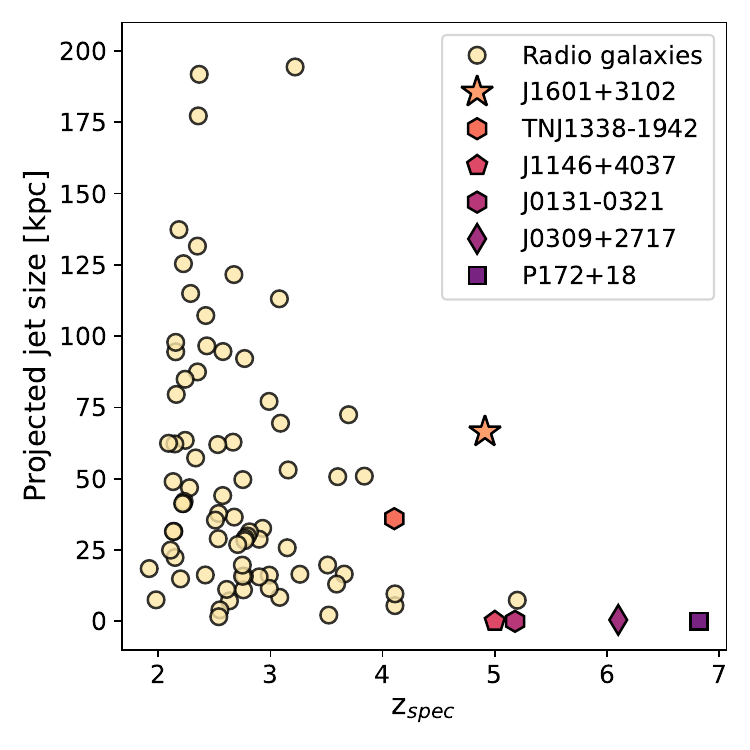}
}\vspace{-0.5cm}
\caption{\label{fig:BH_mass} Physical properties of J1601+3102 compared to literature. \textbf{Left:} Black hole mass of J1601+3102 derived from the Mg\textsc{ii} line versus the bolometric luminosity compared to other known high-$z$ quasars from \cite{Farina2022ApJ...941..106F}. The contours highlight the distribution of SDSS DR16 quasars between $0.27 < z < 2.72$ from \cite{Wu2022ApJS..263...42W}. The systematic error on the BH mass of $\sim$0.55 dex is shown in the bottom right corner. The SMBH mass J1601+3102 is lower than the average of this population, whereas the Eddington ratio is similar. \textbf{Right:} Projected jet size of J1601+3102 compared to known high-redshift radio galaxies \citep{Saxena2019MNRAS.489.5053S} and quasars (J0309+2717; \citealt{Spingola2020A&A...643L..12S}, P172+18; \citealt{Momjian2021AJ....161..207M}, J0131-0321/J1146+4037; \citealt{Shao2022A&A...659A.159S}). J1601+3102 is the first quasar at $z>4$ with a large resolved radio jet.}
\end{figure*}

We derive the black hole properties of J1601+3102 from the fitting routine described in Sect.~\ref{subsec:spectral_fitting}. Specifically, we estimate the BH mass using the FWHM of the Mg\textsc{ii} line and the monochromatic luminosity using the empirical relation derived by \cite{Shen2011ApJS..194...45S} as
\begin{equation}
\label{eq:bh_mass}
    M_{\text{BH}} = 10^{6.74} \times \Bigg(\frac{\text{FWHM}_{\text{Mg}\textsc{ii}}}{1000 \text{~km/s}}\Bigg)^2 \Bigg(\frac{\lambda L_{\lambda,3000}}{10^{44} ~\text{erg/s}}\Bigg)^{0.62},
\end{equation}
where $\lambda$ is the wavelength, $L_{\lambda}$ the monochromatic luminosity at 3000\AA. The uncertainty on the BH mass is predominantly caused by the intrinsic scatter of this relation of $\sim$0.55 dex. Furthermore, we calculate the bolometric luminosity using the relation of \cite{Richards2006ApJS..166..470R} and Eddington luminosity given by 
\begin{align}
\label{eq:lbol}
    L_{\text{bol}} &= 5.15\times\lambda L_{\lambda,3000} \\
    L_{\text{edd}} &= 1.3\times10^{38}\times M_{\text{BH}} .
\end{align}
The results are summarized in Table \ref{tab:table_properties}. We obtain a black hole mass of M$_{\text{BH}}$ = (4.5$^{+1.9}_{-1.2}$)$\times$10$^{8}$ M$_{\odot}$ with a bolometric luminosity of $L_{\text{bol}} = (2.6\pm0.1)\times10^{46}$ erg s$^{-1}$ an Eddington ratio $L_{\text{bol}}/L_{\text{Edd}} = 0.45^{+0.16}_{-0.13}$. We compare these results to the known high-$z$ quasar population in Fig.~\ref{fig:BH_mass}. This indicates J1601+3102 is less massive than the general population, but with a similar accretion rate, meaning J1601+3102 is currently in quasar mode and accreting efficiently. We also show the distribution of quasars between $0.27 < z < 2.72$ with Mg\textsc{ii} line detections from the Sloan Digital Sky Survey (SDSS) Data Release 16 quasar catalog (DR16; \citealt{Wu2022ApJS..263...42W}). To select only quasars with broad emission lines we require FWHM$_{\text{Mg}\textsc{ii}} > 1000$ km s$^{-1}$. For consistency, we calculate their bolometric luminosity and BH mass using equation \ref{eq:bh_mass} and \ref{eq:lbol}. Fig.~\ref{fig:BH_mass} shows that the BH mass of J1601+3102 is comparable to the bulk of the low-$z$ SDSS quasar sample. Lower mass BHs (of $\sim10^{6-8}$ M$_{\odot}$) have also been found at high-$z$ with the discovery of faint AGN with JWST (see e.g. \citealt{Maiolino2023arXiv230801230M, Harikane2023ApJ...959...39H, Matthee2024ApJ...963..129M, Kocevski2024arXiv240403576K}). However, the BH mass of J1601+3102 is still within the scatter of the high-$z$ quasar population and therefore quite ordinary in comparison.

Using these black holes mass and Eddington ratio estimates, we can estimate the expected jet power of J1601+3102 in the thin disk regime, which is given by 
\begin{equation}
    Q_{\text{jet}} = 2 \times 10^{36} \Big(\frac{M_{\text{BH}}}{10^9 M_{\odot}} \Big)^{1.1} \Big(\frac{\lambda_{\text{Edd}}}{0.01} \Big)^{1.2} a^2 \ \text{W}
\end{equation}
with $\lambda_{\text{Edd}}$ the Eddington ratio and $a$ the black hole spin \citep{Meier2002NewAR..46..247M, Orsi2016MNRAS.456.3827O}. To estimate the jet power we assume $a=1$, since radio-loud quasars are thought to have high black hole spin (e.g. \citealt{Maraschi2012JPhCS.355a2016M, Schulze2017ApJ...849....4S}). This calculation yields a jet power estimate of $(8^{+6}_{-4})\times10^{37}$ W or (8$^{+6}_{-4})\times10^{44}$ erg s$^{-1}$. This is on the higher end of the predicted distribution of jet powers by \cite{Saxena2017MNRAS.469.4083S}.

\begin{table}[ht!]
    \centering
    \caption{Multi-wavelength measurements of J1601+3102 from this and previous work.}
    \resizebox{0.98\columnwidth}{!}{
    \begin{tabular}{c c}
    \hline \hline
       RA  &  16:01:49.45  \\
       Dec.  &  +31:02:07.25 \\
       \hline
        & Optical properties \\
       $g, r, z$ & $>24.0$, 23.15$\pm$0.13, 21.19$\pm$0.07 \\
       W1, W2 & 21.51$\pm$0.20, 21.31$\pm$0.35\\ 
       $M_{1450}$ & $-24.75^{+0.31}_{-0.21}$ $^{[1]}$\\
       $\alpha_{\lambda}$ & $-1.71\pm0.05$ \\
       $\lambda L_{3000\AA}$ & (5.0$\pm$0.1)$\times$10$^{45}$ erg s$^{-1}$ \\
       L$_{\text{bol}}$ & (2.6$\pm$0.1)$\times$10$^{46}$ erg s$^{-1}$\\
       $z_{\text{Mg}\textsc{ii}}$ & 4.912$^{+0.004}_{-0.005}$ \\
       FWHM$_{\text{Mg}\textsc{ii}}$ & 2694$^{+510}_{-383}$ km s$^{-1}$ \\
       M$_{\text{BH}}$ & (4.5$^{+1.9}_{-1.2})\times10^{8}$ M$_{\odot}$ \\
       L$_{\text{bol}}$ / L$_{\text{Edd}}$ & 0.45$^{+0.16}_{-0.13}$ \\
       
       \hline
        & Radio properties \\ 
       $R_{2500}$ & $520^{+310}_{-160}$ \\ 
       $R_{4400}$ & $1020^{+720}_{-330}$ \\ 
       $L_{150\text{MHz}, \text{tot}}$ & (2.8$\pm$0.4)$\times10^{28}$ W Hz$^{-1}$ $^{[2]}$\\ 
       $S_{150\text{MHz}, \text{core}}$ & 2.2$\pm$0.3 mJy\\ 
       $S_{150\text{MHz}, \text{north}}$ & 50.6$\pm$5.1 mJy\\ 
       $S_{150\text{MHz}, \text{south}}$ & 10.5$\pm$1.6 mJy\\
       $\alpha_{\text{total}}$ & $-1.27^{+0.06}_{-0.06}$ \\
       $\alpha_{\text{north}}$ & $-1.26^{+0.06}_{-0.06}$ \\
       $\alpha_{\text{south}}$ & $-1.27^{+0.13}_{-0.10}$ \\
       \hline
       \hline \vspace{2pt}
    \end{tabular}}
    \vspace{0.0cm} \raggedright {\small Notes. [1] from \cite{Gloudemans2022A&A...668A..27G} [2] Calculated using the LoTSS-DR2 total flux density of $69.8\pm7.7$ mJy.}
    \label{tab:table_properties}
\end{table}

\section{Discussion}
\label{sec:discussion} 

The large extended radio jet of J1601+3102 is unique because such a potentially $\sim$100 kpc radio jet has never been found above $z>4$, while being common at lower redshift ($z\sim1-2$). The projected jet sizes of J1601+3102 and other known resolved quasars at $z>4$ and radio galaxies at $z>2$ are shown in the right panel of Fig.~\ref{fig:BH_mass}. As discussed, the lack of extended radio sources in the early Universe has previously been attributed to the CMB energy density increasing with a factor (1+$z$)$^4$, causing low-energy photons to be scattered to high energies by the inverse Compton effect. We also potentially see evidence of IC scattering in the LOFAR VLBI image of J1601+3102, since there seems to be a lack of diffuse radio emission between the two radio lobes (see Fig.~\ref{fig:vlbi_im}). Alternatively, it is possible that the diffuse emission may be only revealed at even lower radio frequencies due to the steepness of the radio spectrum. Finally, we note that the high-resolution imaging process could have resolved out extended diffuse emission, however since the source is relatively small and the flux density measurements at low- and high-resolution are similar, this is unlikely to be the main cause of the missing diffuse emission.

Interestingly, the SMBH of J1601+3102 is found to be lower mass compared to the average high-$z$ quasar. Whether or not there is a correlation between the SMBH mass of quasars and their radio-loudness is still being debated in literature with studies both finding significant correlations (e.g. \citealt{McLure2004MNRAS.353L..45M, Seymour2007ApJS..171..353S, Whittam2022MNRAS.516..245W}) and not finding them (e.g. \citealt{Gurkan2019A&A...622A..11G, Macfarlane2021, Arnaudova2024MNRAS.528.4547A}). Another recent study takes a new physically motivated approach to radio-loudness by separating the host galaxy star formation from the AGN contribution to the radio emission and finds that quasars hosting the 20\% most massive SMBHs are 2-3 times more likely to host powerful radio jets than lower mass SMBHs in otherwise similar quasars, however, quasars of all properties can still potentially host luminous jets (Yue et al. subm.). An exceptional Eddington accretion ratio is not strictly needed to generate powerful jets. We conclude that our finding of a relatively low SMBH mass is not in tension with the source exhibiting a huge bright radio jet, especially in the case of a high jet power (see Sect.~\ref{sec:bh_properties}). 

The expected lifetime of the jets $t$ can be approximated from the lobe length $D$, the gas density inside the lobe and the jet power by simplifying Equation A2 of \cite{Kaiser2007MNRAS.381.1548K} to
\begin{equation}
    D = C \Big( \frac{Q_{\text{jet}}}{\rho} \Big)^{1/5} t^{3/5}.
\end{equation}
Assuming a constant gas density of $\rho=10^{-22}$ kg m$^{-3}$ and constant of $C=1.7$ (see \citealt{Kaiser2007MNRAS.381.1548K}) yields an age of $\sim$50 Myr for a lobe size of 66 kpc. However, the physical lobe of J1601+3102 can be as large as 380 kpc for a viewing angle of $\theta = 80^{\circ}$, which would imply an age as high as $\sim$1 Gyr and formation as early as $z\sim10$. Note that the gas density is not expected to be constant around the radio source. Especially in the case of J1601+31, the environment is expected to play a crucial role in boosting the radio emission of the Northern lobe and therefore this age is only a simplified and crude estimate. Improving our understanding of systems like these is crucial to set observational constraints on the formation time of the first radio-loud sources in our Universe.

The average quasar lifetimes are not well constrained with estimated duty cycles of 1 Myr $-$ 1 Gyr (e.g. \citealt{Martini2001ApJ...547...12M, Yu2002MNRAS.335..965Y, Martini2004cbhg.symp..169M, Shen2007AJ....133.2222S}). However, recent works using the HeII proximity zone and quasar clustering suggest quasar accretion episodes of only a few Myrs (e.g. \citealt{Khrykin2021MNRAS.505..649K, Pizzati2024arXiv240312140P}). This suggests J1601+3102 could either be long lived compared to the general quasar population or could show recurrent quasar activity. However, there are no additional hotspots seen in the radio image, which is expected in the case of recurrent activity (e.g. \citealt{Lara1999A&A...348..699L, Nandi2019MNRAS.486.5158N}). 

The discovery in this work shows that these extended radio jets do exist at $z>4$ and we speculate the lack of extended jets at high-$z$ is (at least partly) due to selection effects. J1601+3102 has been discovered by selecting optical dropout sources with low-frequency radio detections (see \citealt{Gloudemans2022A&A...668A..27G}), whereas most radio-loud quasars are discovered at high radio frequencies from VLA observations at 1.4-5 GHz (e.g. \citealt{Wang2007AJ....134..617W, Wang2008ApJ...687..848W, Banados2015ApJ...804..118B, Belladitta2020A&A...635L...7B}). Since extended radio lobes are known to have steep spectral indices and therefore become brighter at low-frequency, it is plausible that many of these extended radio sources have been missed in previous surveys. Furthermore, in previous work high-$z$ candidates have been selected on the basis of having compact morphologies (e.g. \citealt{Saxena2019MNRAS.489.5053S, Knowles2021Galax...9...89K}). Due to the Southern lobe of J1601+3102 not being connected to the Northern component it could easily be mistaken for a compact radio source. However, if the lobes had been slightly less extended, our source might have been classified as extended in the LoTSS-DR2 catalogue (see right panel of Fig.~\ref{fig:vlbi_im}). This discovery therefore demonstrates that quasar and radio galaxy candidates could be incorrectly excluded from high-$z$ searches that enforce the assumption that IC losses keep the extended lobes undetectable.

Finally, we note that the discovery of J1601+3102 demonstrates the existence of SMBHs in the early Universe with extremely efficient or energetic outflows. These could potentially influence the early quenching of galaxies, which have recently been found by JWST observations at similar redshift and point to the very early formation of massive galaxies (e.g. \citealt{Carnall2023Natur.619..716C, Nanayakkara2024NatSR..14.3724N, Looser2024Natur.629...53L}).

\section{Summary}
\label{sec:summary}

In this Letter, we present the discovery of a monster radio jet associated with the extremely radio-loud quasar J1601+3102 at $z\sim5$ from sub-arcsecond resolution imaging at 144 MHz. This radio jet is the largest yet identified at $z>4$ with a size of $>66$ kpc. The 0.3\arcsec\ resolution LOFAR VLBI image shows the radio emission is dominated by a Northern and Southern lobe at a distance of 1.4\arcsec\ (9 kpc) and 8.9\arcsec\ (57 kpc) from the optical quasar, respectively, with steep radio spectral indices of $-1.3$. The Northern lobe is $\sim$5 times brighter than the Southern lobe and $\sim$6 times closer to the nucleus of the quasar. This suggests the Northern lobe is potentially brightened by (extreme) interaction with the surrounding ISM and needs further investigation. 

We detect the Mg\textsc{ii} line in the quasars rest-frame UV spectrum, which gives an estimated black hole mass of (4.5$^{+1.9}_{-1.2}$)$\times$10$^{8}$ M$_{\odot}$ and Eddington accretion ratio of 0.45$^{+0.16}_{-0.13}$ with a bolometric luminosity of L$_{\text{bol}} = (2.6\pm0.1)\times10^{46}$ erg s$^{-1}$. The black hole mass is lower than the general high-$z$ quasar population, demonstrating that an exceptional mass is not strictly necessary to generate a powerful jet in this case. From the jet power, we approximate the expected lifetime of the jets of 50 Myr to $\sim$1 Gyr. This age estimate is highly dependent on the gas density and the viewing angle, which determines the physical jet size. The determination of the viewing angle, by for example X-ray observations, is therefore necessary to pin down the physical size of the radio jet, the advance speed, and the age.

This quasar is unique as it is the first quasar with large extended radio jets in the early Universe, which have remained elusive, potentially due to the increased CMB energy density at high redshift. The LOFAR VLBI image shows potential evidence of IC scattering as well by the lack of diffuse radio emission between the two lobes. This work shows that these large radio jets do exist at $z>4$, but would easily be missed by selecting only at GHz frequencies and requiring compact morphologies. A combination of (blind) spectroscopic confirmation of new high-$z$ radio-loud sources and VLBI observations are necessary to detect more of these monster radio jets in the early Universe and set constraints on the formation time of the first radio-loud sources.

\section*{Acknowledgments}
\noindent
The authors thank Philip Best for his valuable input to the discussion. FS is grateful for the support of STFC [ST/Y004159/1]. E.P.F. is supported by the international Gemini Observatory, a program of NSF NOIRLab, which is managed by the Association of Universities for Research in Astronomy (AURA) under a cooperative agreement with the U.S. National Science Foundation, on behalf of the Gemini partnership of Argentina, Brazil, Canada, Chile, the Republic of Korea, and the United States of America. LKM is grateful for support from the Medical Research Council [MR/T042842/1]. JTS is supported by the Deutsche Forschungsgemeinschaft (DFG, German Research Foundation) - Project number 518006966. KJD acknowledges funding from the STFC through an Ernest Rutherford Fellowship (grant number ST/W003120/1).

This paper is based (in part) on data obtained with the International LOFAR Telescope (ILT) under project codes LC0 015, LC2 024, LC2 038, LC3 008, LC4 008, LC4 034 and LT10 01. LOFAR \citep{vanHaarlem2013A&A...556A...2V} is the Low Frequency Array designed and constructed by ASTRON. It has observing, data processing, and data storage facilities in several countries, which are owned by various parties (each with their own funding sources), and which are collectively operated by the ILT foundation under a joint scientific policy. The ILT resources have benefited from the following recent major funding sources: CNRS-INSU, Observatoire de Paris and Universit\'e d'Orl\'eans, France; BMBF, MIWF-NRW, MPG, Germany; Science Foundation Ireland (SFI), Department of Business, Enterprise and Innovation (DBEI), Ireland; NWO, The Netherlands; The Science and Technology Facilities Council, UK; Ministry of Science and Higher Education, Poland.

Based on observations obtained at the international Gemini Observatory, a program of NSF NOIRLab, which is managed by the Association of Universities for Research in Astronomy (AURA) under a cooperative agreement with the U.S. National Science Foundation on behalf of the Gemini Observatory partnership: the U.S. National Science Foundation (United States), National Research Council (Canada), Agencia Nacional de Investigación y Desarrollo (Chile), Ministerio de Ciencia, Tecnología e Innovación (Argentina), Ministério da Ciência, Tecnologia, Inovações e Comunicações (Brazil), and Korea Astronomy and Space Science Institute (Republic of Korea).

This research made use of \textsc{PypeIt},\footnote{\url{https://pypeit.readthedocs.io/en/latest/}}
a Python package for semi-automated reduction of astronomical slit-based spectroscopy
\citep{Prochaska2020JOSS....5.2308P, Prochaska2020zndo...3743493P}.


\appendix

\section{International LOFAR Telescope calibration procedure}
\label{appendix:LOFAR_ILT}

\subsection{Calibrating systematics}
The systematic corrections were obtained from the flux density calibrator observation, using the LINC Calibrator pipeline. This finds and corrects for three effects in High Band Antenna observations. First an average time-independent offset between the XX and YY correlations is taken out, as for an unpolarised source no difference between the two is expected. Secondly, station bandpasses to convert correlator units to physical units are derived. Finally, an average clock offsets between all stations and a reference station close to the centre of the array is taken out. The remote and international stations have their own independent atomic clocks, which can drift approximately $20\ \mathrm{ns}$ per $20\ \mathrm{min}$ and thus are periodically synchronised to GPS \citep{vanHaarlem2013A&A...556A...2V}.

\subsection{Direction independent calibration of Dutch stations}
The target observation is processed using the LINC Target pipeline, with the recently added self-calibration functionality. This first applies the corrections found by LINC Calibrator and corrects for ionospheric Faraday rotation using RMextract\footnote{https://github.com/lofar-astron/RMextract} \citep{Mevius2018}. Finally, the data is concatenated into a single frequency band, after which a phase-only calibration against a model from the Tata Institute of Fundamental Research (TIFR) Giant Metrewave Radio Telescope (GMRT) Sky Survey (TGSS) Alternative Data Release 1 (ADR1) is performed, in order to correct for direction-independent ionospheric effects. At HBA frequencies this is mainly a propagation delay \citep{deGasperin2019A&A...622A...5D}. DP3's ``smoothness constraint'' was used to constrain the phase solutions to smooth behaviour in the frequency domain over a running $5\ \mathrm{MHz}$ window. This was done to alleviate potential adverse effects from calibrating against the TGSS ADR1 model which is less sensitive and has lower angular resolution than the ILT observations. An inner $uv$ cut of $200\lambda$ was applied to match that of TGSS ADR1. Next, it makes an image of the field at $\sim 6''$ angular resolution using \texttt{WSClean}, and does one round of phase-only self-calibration against the model obtained from that image.

\subsection{International station calibration}
Calibration of the international stations was done in three steps: calibration on a reference calibration source, self-calibration on the target source and finally an astrometric correction on a nearby compact source with an identified optical counterpart.

Following the usual strategy (see e.g. \citealt{Morabito2022A&A...658A...1M, Sweijen2022NatAs...6..350S}), the international baselines were calibrated using a calibrator source in the Long Baseline Calibrator Survey (LBCS; \citealt{Moldon2015A&A...574A..73M, Jackson2016A&A...595A..86J,Jackson2022A&A...658A...2J}), also referred to as a ``delay calibrator". Criteria for a good calibrator candidate are, for example, high flux density and compactness. For P240+30 the selected calibrator was ILTJ155955.03+304223.7; a fairly compact source with a flux density, as measured in LoTSS, of $S_\mathrm{LoTSS} = 1.0$ Jy, making it a prime calibration candidate. The visibilities were phase-shifted to the location of this source, and subsequently averaged to a time resolution of $8\ \mathrm{s}$ and a frequency resolution of $390.625\ \mathrm{kHz}$. Delay calibration was performed through self-calibration on this phase-shifted set of visibilities.

Due to the lack of an appropriate starting model, the self-calibration was started assuming a point source at the phase centre of unit flux density. First, the data was converted from linear correlations to circular correlations. This makes correcting for Faraday rotation easier, as it will manifest as a phase difference between the parallel-hand RR and LL correlations, instead of moving signal into the cross-hand XY and YX correlations. The self-calibration procedure consisted of multiple cycles and several perturbations within each cycle. In the first four cycles phase corrections were derived. From the fifth cycle on, when self-calibration had started to converge, amplitude corrections were also allowed. To reduce the impact of emission still seen on shorter baselines, an inner cut of $20\ \mathrm{k}\lambda$ was used for the calibration (N.B. this was only used in the \textit{calibration} aspect. No uv cut was used during imaging). In summary, the following perturbations in facetselfcal were used:

\begin{enumerate}
    \item \textbf{Faraday rotation} - using the \texttt{scalarphasediff} perturbation, a phase difference $\Delta\phi_\mathrm{FR} = \phi_\mathrm{RR} - \phi_\mathrm{LL}$ was derived to capture the effects of Faraday rotation. This assumes that the source is not circularly polarised.
    \item \textbf{Fast phases on international stations} - using the \texttt{scalarphase} perturbation, polarisation-independent phase correction were derived for the international stations. Dutch station solutions were reset to 0 phase (i.e. no correction), as those will be derived in the following perturbations.
    \item \textbf{Residual fast phases on remote stations} - using the \texttt{scalarphase} perturbation, polarisation-independent phase corrections were derived for the remote stations. Solutions for the core stations were reset to 0 phase, to be corrected in the next perturbation.
    \item \textbf{Residual phases on core stations} - using the \texttt{scalarphase} perturbation, polarisation-independent phase corrections were derived for the core stations.
    \item \textbf{Amplitude corrections} - using the \texttt{scalarcomplexgain} perturbation, amplitude corrections were derived. These capture e.g. residual bandpass correction or errors in the primary beam model.
    \item \textbf{Polarisation corrections} - using the \texttt{fulljones} perturbation, full-polarisation corrections were derived for all four correlations. This corrects for leakage between the correlations, under the assumption that the source is unpolarised.
\end{enumerate}

In total, eight cycles of self-calibration were done using perturbations $1-5$. After each calibration cycle, an image was made using a pixel size of $0.075\arcsec/\mathrm{pix}$, using a robust weighting of $-1.5$.

\subsection{Flux density scale}
\begin{figure}
    \centering
    \includegraphics[width=\columnwidth, trim={1cm 7cm 1cm 7cm}, clip]{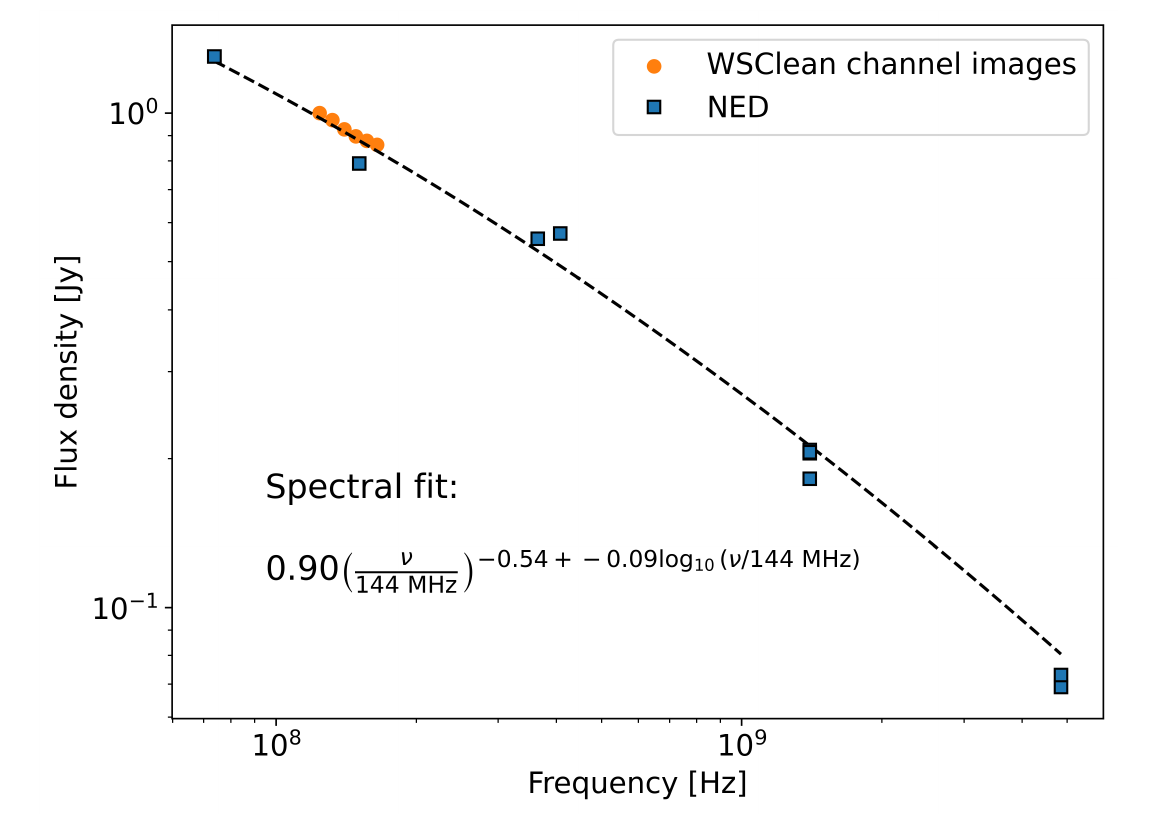}
    \caption{The fitted spectrum (dashed line) of the delay calibrator ILTJ155955.03+304223.7, based on the photometry available in NED (squares) and the flux density as measured from the channel images output by \texttt{WSClean} (circles).}
    \label{fig:spec_inband}
\end{figure}

Amplitude solutions were normalised during self-calibration, preventing a major drift in the flux density scale. However, the transfer of the bandpasses from the calibrator observation to the target field can still contain errors. Therefore, once self-calibration had converged, the CLEAN-component model was rescaled to the expected flux density and spectral index. The spectrum of the source was fitted using archival data points from the NASA Extragalactic Database (NED\footnote{https://ned.ipac.caltech.edu/}). These were used to fit a flux density $S_0$, a spectral index $\alpha_1$ and a spectral curvature $\alpha_2$ using the form
\begin{equation}
    S_\nu = S_0 \left(\frac{\nu}{\nu_\mathrm{ref}}\right)^{\alpha_1 + \alpha_2 \log_{10}\nu / \nu_{\mathrm{ref}}}
\end{equation}
where $\nu_\mathrm{ref}$ is the reference frequency for which to derive the parameters. For our case this was set to $144\ \mathrm{MHz}$. The self-calibration of the delay calibrator was then repeated using the updated model as a starting model, except now all perturbations were solved only once, directly after each other and a polarisation correction was added to correct for leakage. The latter assumes that this source is unpolarised, which we base on the fact that it is not reported in the LoTSS DR2 polarised source catalogue of \cite{O'Sullivan2022yCat..74952607O}. Figure~\ref{fig:spec_inband} shows the fitted spectrum based on the photometry available in NED and the flux density as measured from the channel images output by \texttt{WSClean}. The flux density was measured in each channel by summing pixels within a $>5\ \mathrm{sigma}_\mathrm{rms}$ region of the Stokes-I image and dividing by the appropriate beam area in pixels.

\begin{table}
    \centering
    \caption{Delay calibrator self-calibration parameters}
    \resizebox{\columnwidth}{!}{
    \begin{tabular}{l l l}
        \hline \hline
        Perturbation & Solution & Smoothness  \\
         & interval & constraint [MHz] \\
        \hline
        scalarphasediff & $4$ min & $10.0$ \\
        scalarphase & $8$ s & $1.25$ \\
        scalarphase & $8$ s & $10.0$ \\
        scalarphase & $60$ s & $10.0$ \\
        scalarcomplexgain & $1800$ s & $10.0$ \\
        fulljones & $1800$ & $1.0$ \\
        \hline
    \end{tabular}}
    \label{tab:self_calibration_params}
\end{table}

\subsection{Target self-calibration and astrometry}
A new phase-shifted data set was created for the J1601+3102, to which the final calibration solutions derived using the delay calibrator were applied. Notable direction-dependent effects remain, as the target and delay calibrator are separated by $0.5^\circ$ the sky. These effects are dominated by ionospheric perturbations. Self-calibration on the target therefore provide a measure of the \texttt{tec} perturbation, which exploits the ionospheric pertubration's functional form $\phi \propto \mathrm{dTEC} / \nu$ to help constrain the solutions. The solution interval was calculated dynamically during the self-calibration cycles based on the signal detected on baselines $\gtrsim 148\ \mathrm{km}$. This gave a solution interval of approximately $\sim 10\ \mathrm{min}$.

An astrometric correction was derived using another compact radio source, ILTJ160147.25+310222.4 which was near the main target and for which an optical counterpart had been identified in LoTSS-DR2. A compact source is preferred to reduce the ambiguity of determining the "centre" of a source. CASA's \texttt{imfit} task was used to fit a two-dimensional Gaussian profile to this compact radio source in the sub-arcsecond resolution ILT image. Using the centroid coordinates, the image was then shifted to match the optical coordinates. This correction was $0.92\arcsec$.

\section{(Near-)infrared data reduction}
\label{appendix:NIR_data_reduction}

\begin{figure*}
    \centering
    \includegraphics[width=0.9\linewidth]{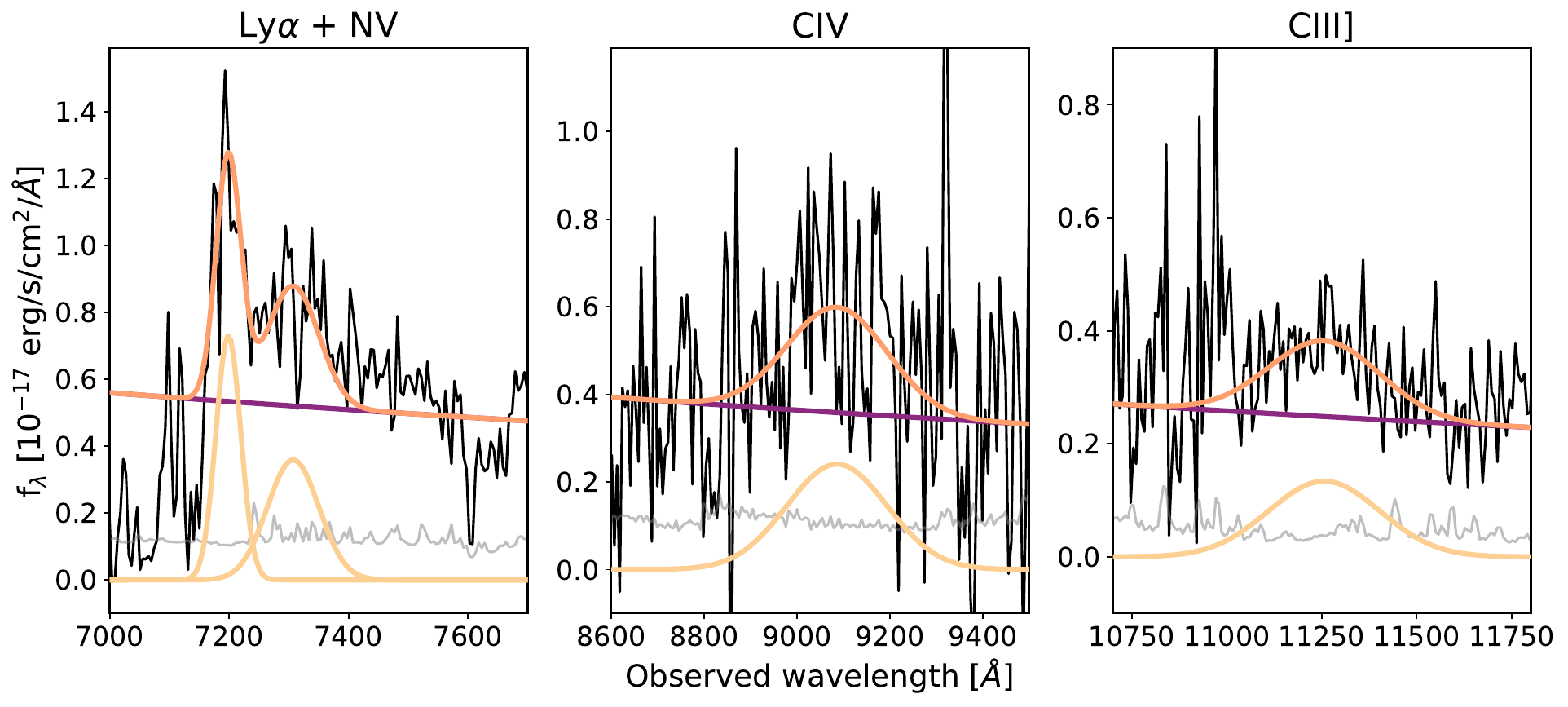}
    \caption{Zoom-in on the Ly$\alpha$ and N$\textsc{v}$ emission lines (left), C$\textsc{iv}$ line (middle), and C$\textsc{iii]}$ line (right). The spectrum is fitted using a powerlaw and iron continuum (purple) and single Gaussian emission lines (yellow). The noise spectrum is indicated in grey.}
    \label{fig:emission_lines}
\end{figure*}

We reduced the data using the \textsc{python} package Python Spectroscopic Data Reduction Pipeline (\textsc{PypeIt}\footnote{\url{https://github.com/pypeit/PypeIt}}; \citealt{Prochaska2020JOSS....5.2308P}), which provides semi-automated reduction for spectroscopic data. The pipeline performed basic image processing on all 2D single exposures such as flat fielding and cosmic ray identification and masking (using the L.A. Cosmic Ray rejection algorithm; \citealt{vanDokkum2001PASP..113.1420V}). The images are then sky-subtracted using difference imaging between the AB dithered pairs and a 2D BSpline fitting procedure, and wavelength-calibrated using the night sky lines. The 1D spectrum is automatically extracted from the 2D images using the optimal spectrum extraction procedure of \cite{Horne1986PASP...98..609H}. Before co-adding the individual 1D spectra, the spectra are flux-calibrated using the standard star. The stacked spectrum is corrected for telluric absorption by fitting a model based on grids from the Line-By-Line Radiative Transfer Model (LBLRTM4; \citealt{Clough2005JQSRT..91..233C, Gullikson2014AJ....148...53G}). 

\subsection{Spectral fitting}
\label{appendix:spectral_fitting}

As discussed in Sect.~\ref{subsec:spectral_fitting}, the Ly$\alpha$, N\textsc{v}, C\textsc{iv}, and C$\textsc{iii]}$ emission lines are fitted with a single Gaussian after subtracting the continuum model, consisting of a powerlaw and iron pseudocontinuum. A zoom-in on these emission lines and their best fits are shown in Fig.~\ref{appendix:spectral_fitting}. The resampling procedure (as outlined in in Sect.~\ref{subsec:spectral_fitting}), results in redshift values of $z_{\text{Ly}\alpha} = 4.920\pm0.002$ and $z_{\text{CIV}} = 4.866\pm0.008$ for the Ly$\alpha$ and C\textsc{iv} line, respectively. The C\textsc{iv} line of quasars is known to be often blueshifted with respect to the Mg\textsc{ii} line due to outflows (e.g. \citealt{Gaskell1982ApJ...263...79G}). Our C\textsc{iv} and Mg\textsc{ii} line fits result in a measured blueshift of $\Delta v_{\text{MgII} - \text{CIV}}=$2350$\pm$450 km s$^{-1}$, which is close to the median value of $\Delta v_{\text{MgII} - \text{CIV}}\sim$1850 km s$^{-1}$ found in literature for quasars at $z\sim6$ \citep{Schindler2020ApJ...905...51S}.

\bibliography{main}{}
\bibliographystyle{aasjournal}



\end{document}